\documentclass[bibyear]{aa} 

\usepackage{graphicx}
\usepackage{amsmath}
\usepackage{float} 
\usepackage{caption}

\usepackage{txfonts}

\begin{document}


\title{Power-2 limb-darkening coefficients for the $uvby$, $UBVRIJHK$, SDSS $ugriz$, \textit{Gaia}, \textit{Kepler,} and TESS photometric systems 
        \thanks{\bf Tables 1-3 are only available at the CDS
                via anonymous ftp to cdsarc.u-strasbg.fr (130.79.128.5) or via
                http://cdsarc.u-strasbg.fr/viz-bin/cat/J/A+A/XXX/XXX }}

\subtitle{I. ATLAS stellar atmosphere models}

\titlerunning {Limb-darkening coefficients for the power-2 law}

\author{A. Claret~\inst{1,2}  \and J. Southworth \inst{3}}


\institute{Instituto de Astrof\'{\i}sica de Andaluc\'{\i}a, CSIC, Apartado 3004, 18080 Granada, Spain
      \and Dept.\ F\'{\i}sica Te\'{o}rica y del Cosmos, Universidad de Granada, Campus de Fuentenueva s/n,  10871, Granada, Spain
      \and Astrophysics Group, Keele University, Staffordshire, ST5 5BG, UK
}

\date{Received; accepted; }

\abstract
   {Limb darkening is an important stellar phenomenon and must be accounted for in the study of stellar spectra, eclipsing binaries, transiting planetary systems, and microlensing events.
    The power-2 limb-darkening law provides a good match to the specific intensities predicted by stellar atmosphere models: it is better than other two-parameter laws and is only surpassed by the four-parameter law.}
   {Predictions of the limb-darkening coefficients for the power-2 law are not widely available. We therefore compute them, using stellar atmosphere models generated by the ATLAS (plane-parallel) code.}
   {Limb-darkening coefficients were computed for the space missions \textit{Gaia}, \textit{Kepler}, and TESS as well as for the photometric systems $uvby$, $UBVRIJHK$, and SDSS $ugriz$.
    The calculations were performed by adopting the Levenberg-Marquardt least-squares minimisation method and were computed with a resolution of 100 equally spaced viewing angles.
    We used 9586 model atmospheres covering 19 metallicities, effective temperatures of 3500 to 50000\,K, $\log g$ values from 0.0 to 5.0, and microturbulent velocities of 0, 1, 2, 4, and 8\,km\,s$^{-1}$.}
   {We confirm the superiority of the power-2 law, in terms of the quality of the fits, over other two-parameter laws. This is particularly relevant for the quadratic law, which is widely used.}
{We recommend  the use of the power-2 law  in cases where a two-parameter law  is needed.}


\maketitle


\section{Introduction}

The phenomenon of limb darkening (LD) is the apparent decrease in  intensity of the surface of a star from the centre of its disc to the limb. The effect is not intrinsic to  the star, but arises due to the viewing geometry of the observer. Viewing angles normal to the  stellar surface experience photons emitted from deeper, and thus hotter, layers of the atmosphere, whereas the atmospheric opacity causes slanted viewing angles to `see' only outer, and thus cooler, layers of the atmosphere.

The Sun has for centuries been known to show the phenomenon of LD (Bouguer 1760). It was first parameterised by the linear `law' (Schwarzschild 1906), which permits only a constant gradient in the decrease in surface brightness as a function of distance from the centre of the stellar disc:
\begin{equation}
\frac{I(\mu)}{I(\mu=1)}=  1 - u(1 - \mu)
,\end{equation}
where $I$ is the specific intensity of light as a function of position on the stellar disc, $u$ is the linear limb-darkening coefficient (LDC), and $\mu = \cos\gamma$, where $\gamma$ is the angle between the observer's line of sight and the surface normal.

The linear LD law is too simplistic, so many other LD parameterisations have been proposed. Some of the most widely used are the quadratic law (Kopal 1950),
\begin{equation}
\frac{I(\mu)}{I(\mu=1)}=  1 - a(1 - \mu) - b(1-\mu)^2,
\end{equation}
 the square-root law (D{\'{\i}}az-Cordov\'es \& Gim\'enez 1992),
\begin{equation}
\frac{I(\mu)}{I(\mu=1)}=  1 - c(1 - \mu) - d(1-\sqrt{\mu}),
\end{equation}
and the logarithmic law (Klinglesmith \& Sobieski 1970),
\begin{equation}
\frac{I(\mu)}{I(\mu=1)}=  1 - e(1 - \mu) - f~\mu~\log(\mu),
\end{equation}
where $a$, $c$, and $e$ are the linear LDCs, $b$ is the quadratic LDC, $d$ is the square-root LDC, and $f$ is the logarithmic LDC. The two-parameter laws provide an improved representation of reality as they allow for a slow drop-off of brightness near the centre of the stellar disc and a faster decrease in the region of the limb.

In addition to these, Hestroffer (1997) introduced a two-parameter law that involves a power of $\mu$, which has become known as the power-2 law:
\begin{equation}
\frac{I(\mu)}{I(\mu=1)}=  1 - g(1 - \mu^h)
,\end{equation}
where $g$ and $h$ are the corresponding LDCs. Here, it can be seen that the coefficient $g$ gives the specific  normalised intensity at the limb of the star ($\mu = 0$) and $h$ governs how much the specific intensity departs from the linear law. The power-2 law has been found to fit specific intensity profiles very well (Morello et al.\ 2017; Maxted 2018) and has also been implemented in an efficient algorithm for planetary transits (Maxted \& Gill 2019). Short et al.\ (2019) have shown that the power-2 law provides a better match to the specific intensities for spherical stellar atmosphere models than other two-parameter laws. To our knowledge, the only known LD law that surpasses it in quality of fit is the four-parameter law (Claret 2000).

 Limb darkening is a vital phenomenon in several areas of astrophysics where a stellar disc is resolved in some way, including high-precision spectroscopy, eclipsing binary star systems, transiting extrasolar planets, interferometry, and stellar microlensing events. The power-2 law is beginning to be widely used in these areas.

Theoretical LDCs for the power-2 law have been presented for a subset of situations: DA, DB, and DBA white dwarfs (Claret et al. 2020), solar-type stars (Maxted 2018), and the passband of the CHEOPS space mission (Claret 2021). However, theoretical LDCs covering a wide range of stellar properties and passbands have not so far been available.

The aim of the current work is to provide users with LDCs for the power-2 law covering a wide range of stellar types and passbands (see Claret 2000, 2011). The passbands are those for the Str\"omgren $uvby$, Johnson-Cousins $UBVRIJHK$, and SDSS $ugriz$ photometric systems, plus those for the \textit{Gaia}, \textit{Kepler,} and TESS space telescopes. In Sect. 2 we briefly describe the details of the computational method for deriving the LDCs adopting the ATLAS models. Sections 3 and 4 present the results of our calculations plus comparisons to the other two-parameter laws. We conclude in Sect. 5.


\section{Numerical method}

We adopted the least-squares method (LSM), adapted to the non-linear case, to compute the LDCs. Before applying the LSM  to each passband, the specific intensities for the $uvby$, $UBVRIJHK$, $ugriz$, \textit{Gaia}, \textit{Kepler,} and TESS photometric systems were integrated using the following equation:
\begin{equation}
I_{a}(\mu) = h^{-1} {\int_{\lambda_1}^{\lambda_2} { I(\lambda,\mu) S(\lambda)\lambda^{-1} d\,\lambda}\over\int_{\lambda_1}^{\lambda_2} { S(\lambda)  d\,\lambda}},
\end{equation}
where $h$ is Planck's constant, $\lambda$ is the wavelength, $I_{a}(\mu)$ is the specific intensity in passband $a$, $I(\lambda,\mu)$ is the monochromatic specific intensity, and $S(\lambda)$ is the response function. For the $uvby$, $UBVRIJHK,$ and $ugriz$ passbands, the function also takes into account the transmission of one airmass of the Earth's atmosphere  response. The passbands used were obtained from Spanish Virtual Observatory Filter Profile Service\footnote{\texttt{http://svo2.cab.inta-csic.es/theory/fps/}}, except for the cases of Str\"omgren $uvby$, the passband of which was obtained from the Observatorio de Sierra Nevada, Granada, Spain (C.\ C\'ardenas, private communication),  and  $JHK,$ obtained from the Observatorio del Teide-IAC, Spain (see also  Alonso et al. 1994). 

The LDC calculations were performed adopting a total of 9586 ATLAS stellar model atmospheres\footnote{\texttt{http://kurucz.harvard.edu/grids.html}}. These covered 19 metallicities [M/H] (from 10$^{-5}$ to 10$^{+1}$ times the solar abundance), surface gravities $\log g$ from 0.0 to 5.0, effective temperatures $T_{\rm eff}$ from 3500\,K to 50000\,K, and five microturbulent velocities ($V_\xi = 0$, 1, 2, 4, 8\,km\,s$^{-1}$). In each case we used a total of 100 $\mu$ points equally spaced in $\mu$, rather than the 17 points normally adopted in the ATLAS models.

The corresponding merit function to show the quality of the fit to the $I(\mu)$ values can be written as
\begin{equation}
{\chi^2} = \sum_{i=1}^{N} \left( {y_i - Y_i}\right)^2
,\end{equation}
where $y_i$ is the model intensity at point $i$, $Y_i$ is the fitted function at the same point, and $N$ is the number of $\mu$ points.


\begin{figure*}
        \includegraphics[height=18.cm,width=13cm,angle=-90]{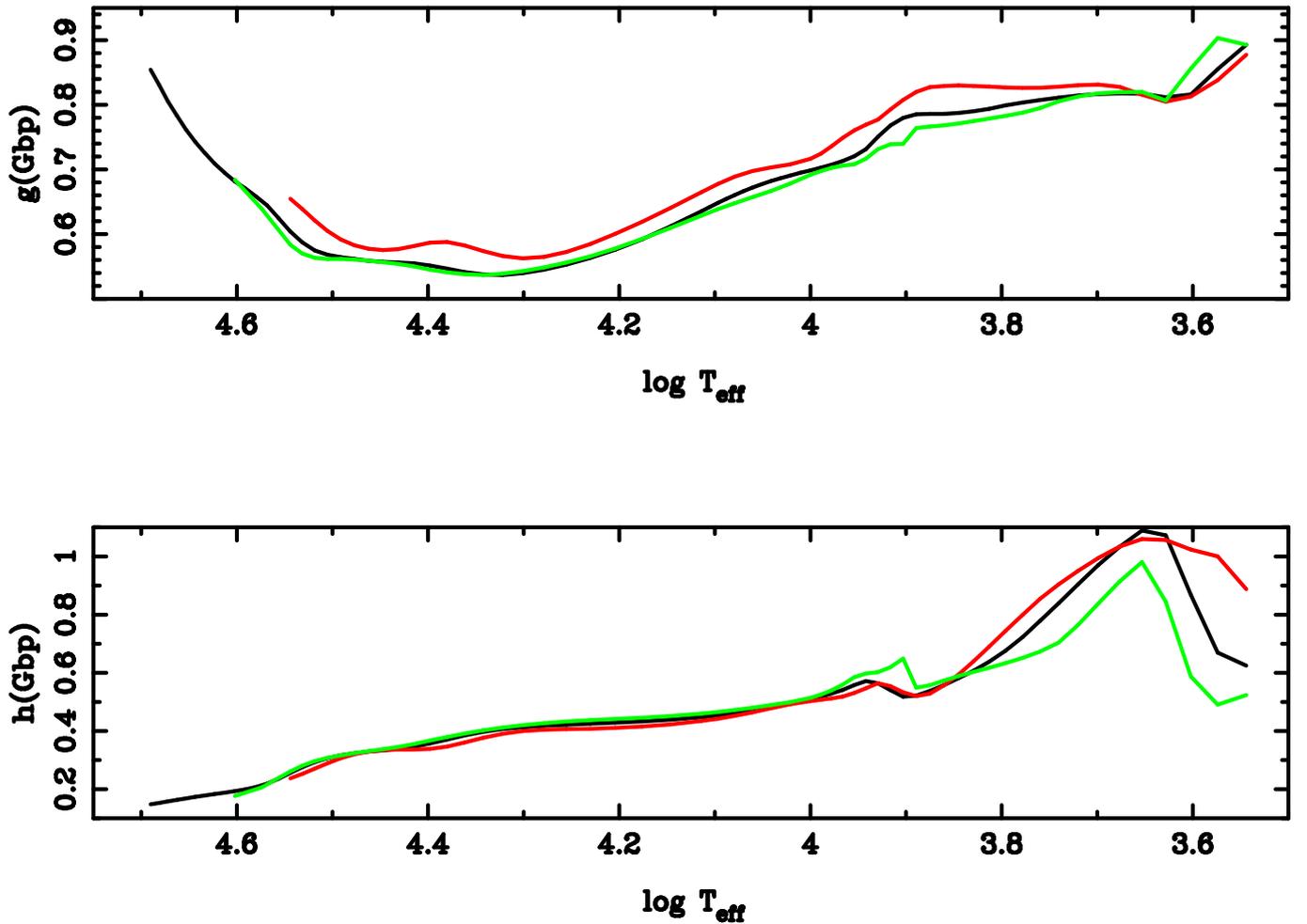}
        \caption{\label{fig:gh} Coefficients   $g$ and $h$  for the filter Gbp of Gaia as a function of $T_{\rm eff}$ and [M/H].
        Black lines represent [M/H] $= 0.0$, red lines [M/H] $= 1.0$, and green
        lines [M/H] $= -1.0$. In all cases, $\log g = 4.5$ and $V_\xi = 2$~km~s$^{-1}$.}
\end{figure*}

\begin{figure}
        \includegraphics[height=8.9cm,width=6cm,angle=-90]{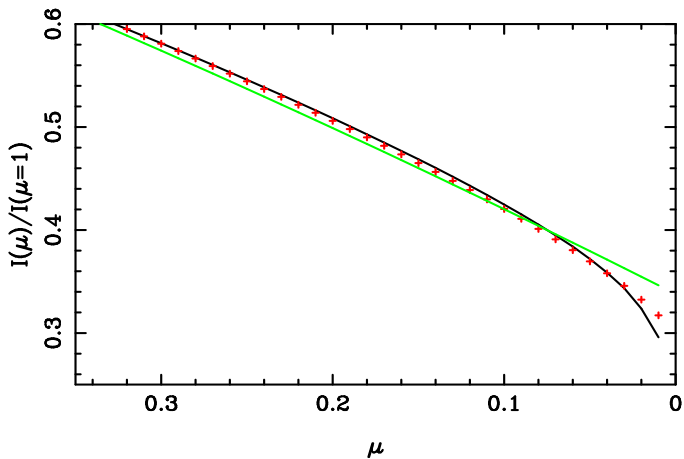}
        \caption{\label{fig:4500k} Angular distribution of the specific intensity for a model with
        $T_{\rm eff} = 4500$\,K, $\log g = 5.0$, [M/H] $=$ 0.0, and $V_\xi = 2$\,km\,s$^{-1}$
        for the TESS passband (black continuous line). Red crosses denote the fitting adopting
        the power-2 law, and the green line represents the quadratic LD law approach.}
\end{figure}

\begin{figure}
        \includegraphics[height=8.9cm,width=6cm,angle=-90]{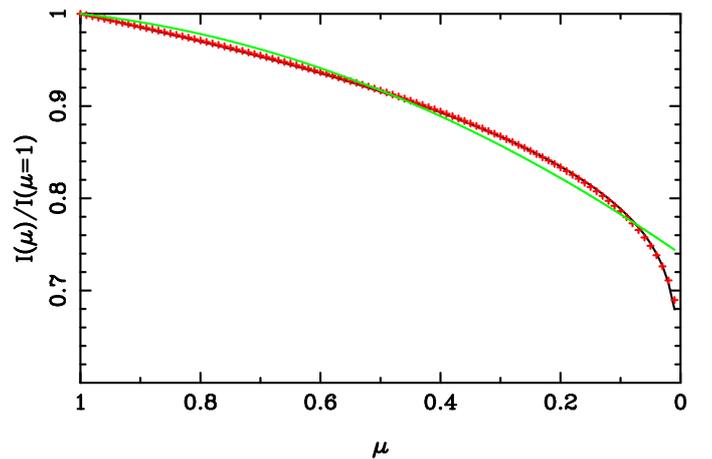}
        \caption{\label{fig:23000k} Same as Fig.~\ref{fig:4500k} but for the model with $T_{\rm eff} = 23000$\,K,
        $\log g = 5.0$, [M/H] $= 0.0,$ and $V_\xi = 2$\,km\,s$^{-1}$.}
\end{figure}

\begin{figure}
        \includegraphics[height=8.9cm,width=6cm,angle=-90]{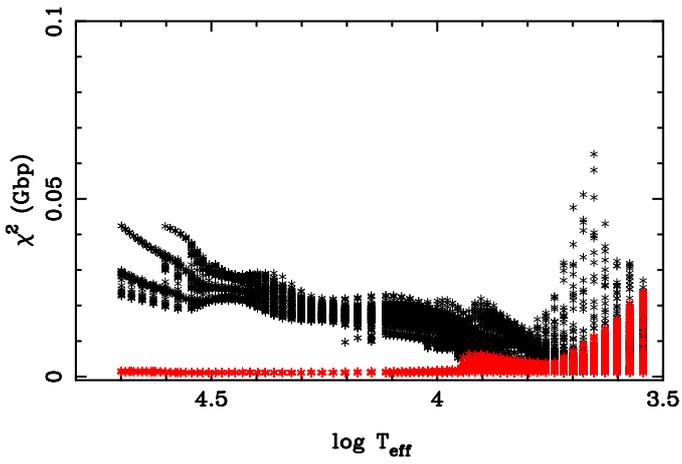}
        \caption{\label{fig:pvq:gbp} Merit function for the ATLAS models as a function of $\log T_{\rm eff}$
        for the \textit{Gaia} $G_{\rm BP}$ passband. The power-2 law is shown using red asterisks
        and the quadratic law using black asterisks. The results for all models are shown.}
\end{figure}

\begin{figure}
        \includegraphics[height=8.9cm,width=6cm,angle=-90]{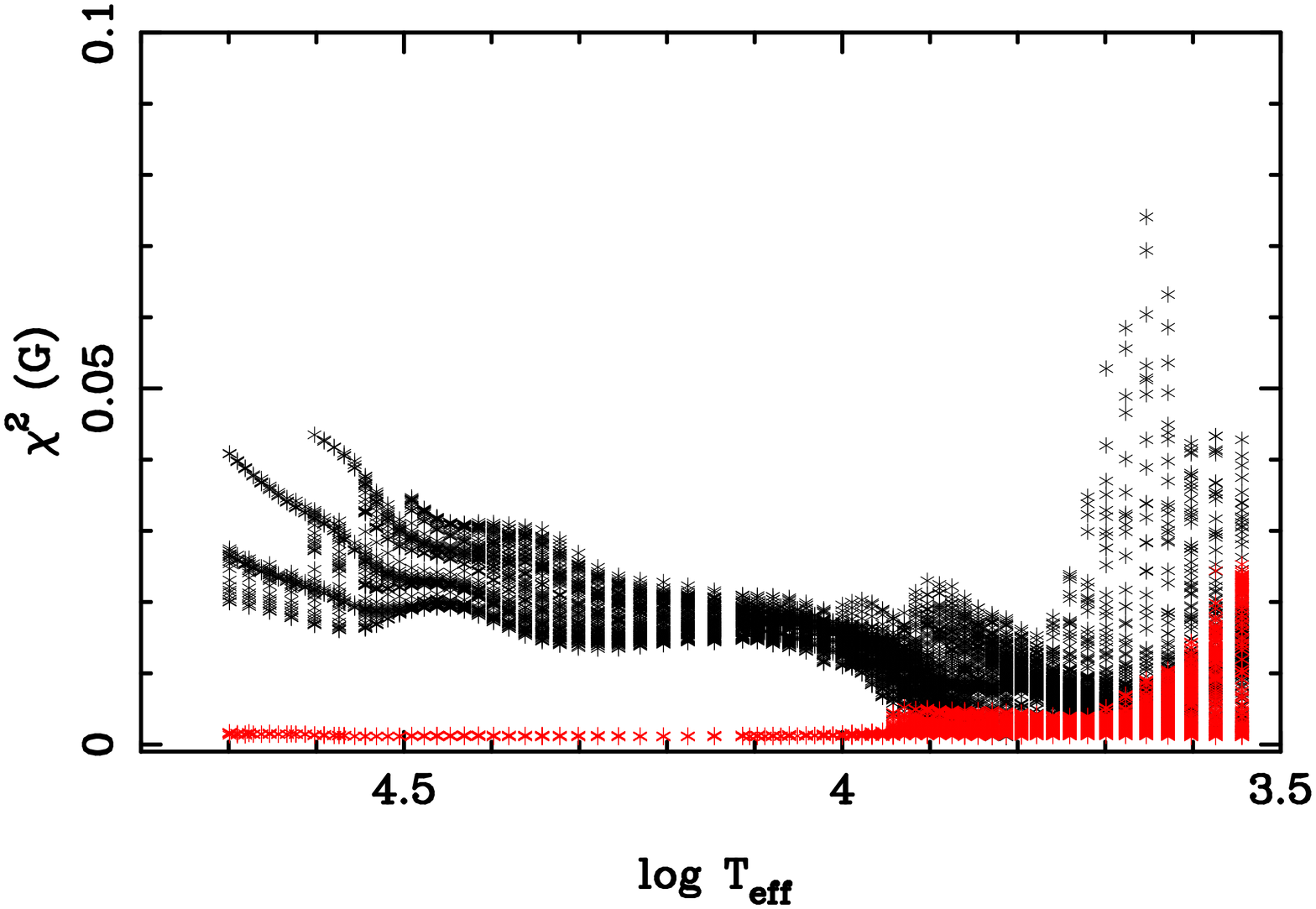}
        \caption{\label{fig:pvq:g} Same as Fig.~\ref{fig:pvq:gbp} but for the \textit{Gaia} $G$ passband.}
\end{figure}

\begin{figure}
        \includegraphics[height=8.9cm,width=6cm,angle=-90]{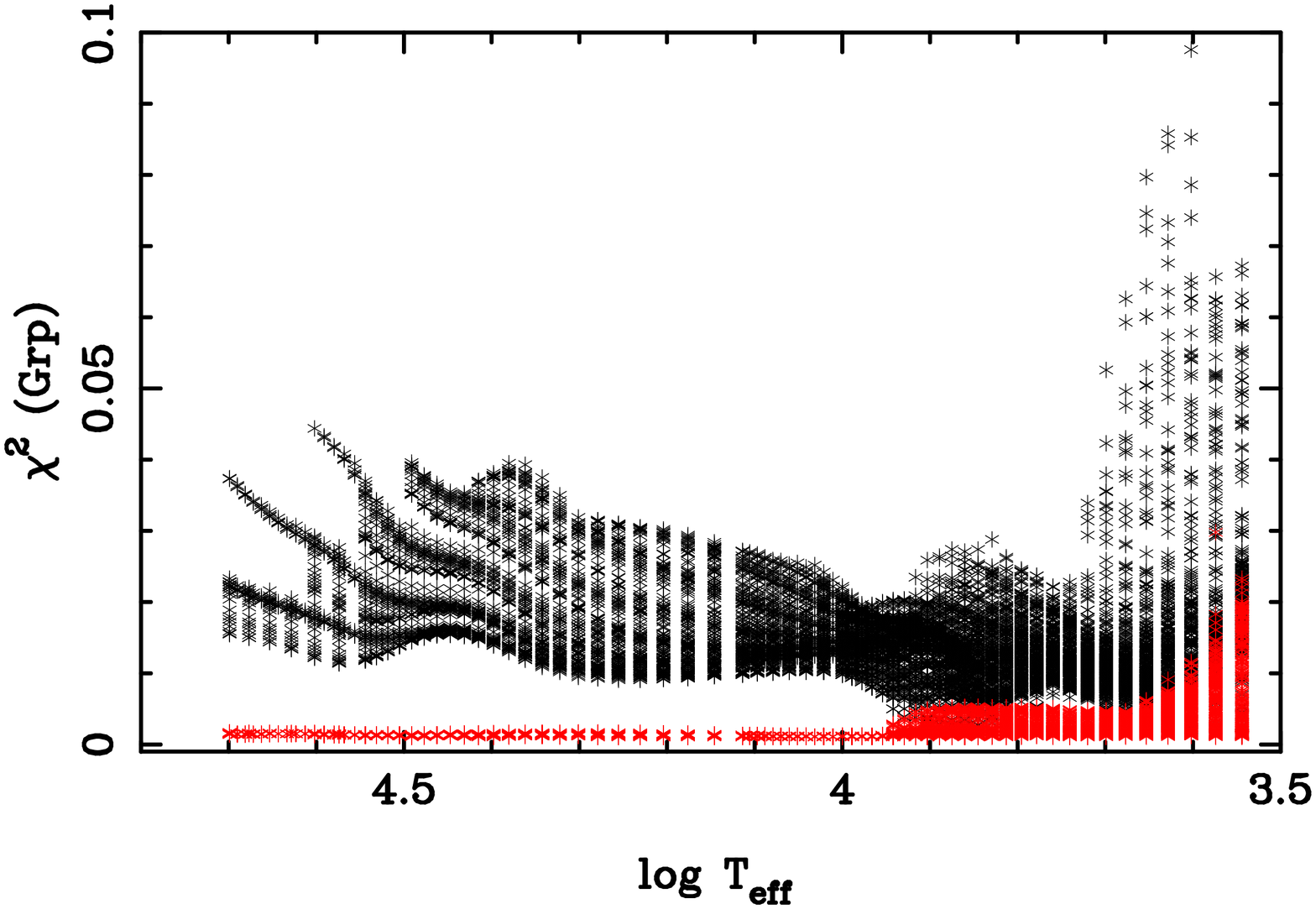}
        \caption{\label{fig:pvq:grp} Same as Fig.~\ref{fig:pvq:gbp} but for the \textit{Gaia} $G_{\rm RP}$ passband.}
\end{figure}

\begin{figure}
        \includegraphics[height=8.9cm,width=6cm,angle=-90]{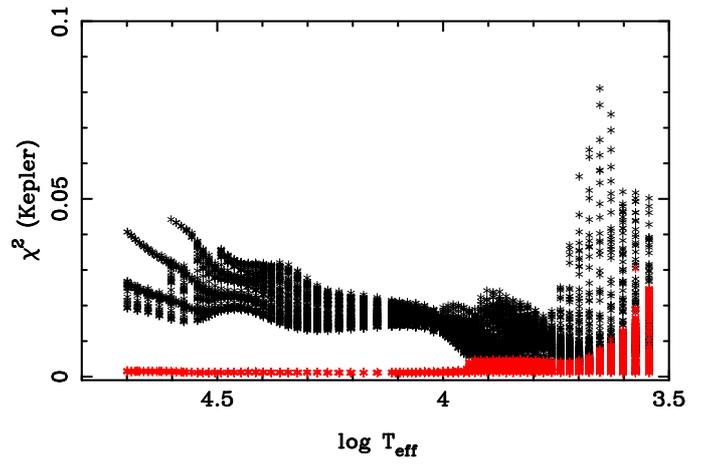}
        \caption{\label{fig:pvq:kep} Same as Fig.~\ref{fig:pvq:gbp} but for the \textit{Kepler} passband.}
\end{figure}

\begin{figure}
        \includegraphics[height=8.9cm,width=6cm,angle=-90]{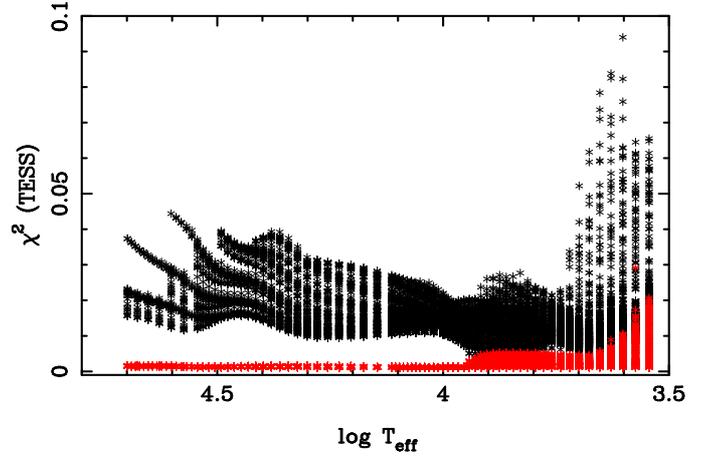}
        \caption{\label{fig:pvq:tess} Same as Fig.~\ref{fig:pvq:gbp} but for the TESS passband.}
\end{figure}

\begin{figure}
        \includegraphics[height=8.9cm,width=6cm,angle=-90]{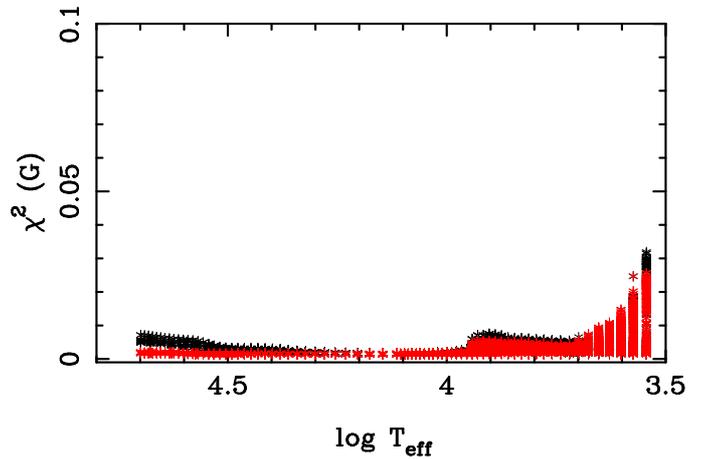}
        \caption{\label{fig:pvs:g} Same as Fig.~\ref{fig:pvq:g} but for the square-root law (black asterisks) versus the power-2 law (red asterisks).}
\end{figure}

\begin{figure}
        \includegraphics[height=8.9cm,width=6cm,angle=-90]{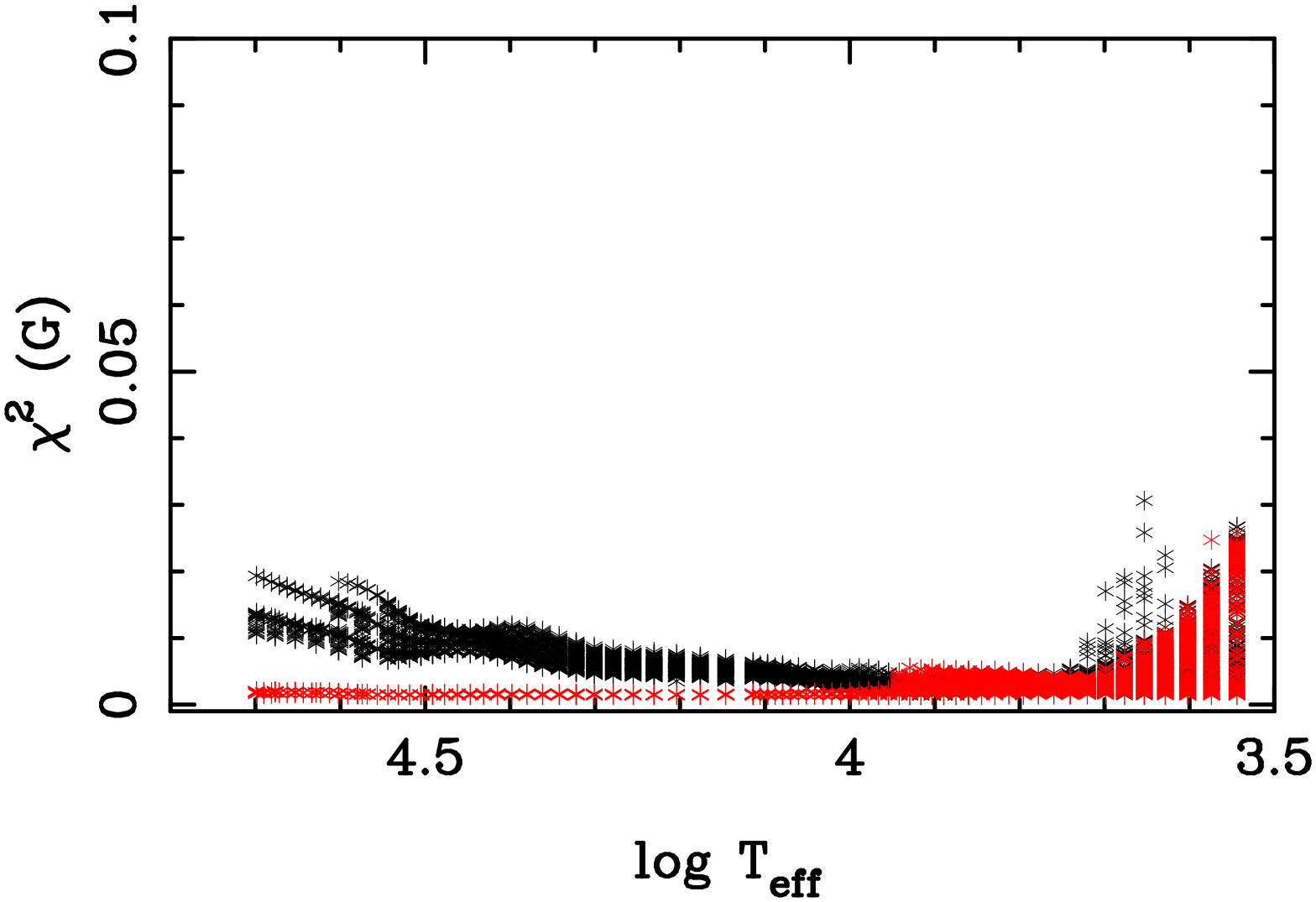}
        \caption{\label{fig:pvl:g}Same as Fig.~\ref{fig:pvq:g} but for the logarithmic law (black asterisks) versus the power-2 law (red asterisks).}
\end{figure}


\section{Results and comparison with the quadratic law}

The power-2 LDCs calculated in this work are made available in three tables (online only; see Table A.1). Table 1 contains the LDCs and $\chi^2$ values for the \textit{Gaia}, \textit{Kepler,} and TESS passbands, Table 2 contains this information for the SDSS $ugriz$ bands, and Table 3 contains this information for the $uvby$ and $UBVRIJHK$ bands. The LDCs for the CHEOPS mission, computed using the ATLAS and PHOENIX models, can be found in Claret (2021).

Figure~\ref{fig:gh} illustrates the behaviour of the coefficients $g$ and $h$ for a series of models with differing $T_{\rm eff}$ and [M/H] values. It can be seen that the coefficient $h$ does not depend very strongly on metallicity, except at low $T_{\rm eff}$. The change in slope, around $\log T_{\rm eff} = 3.9$, is due to the onset of convection. It can also be seen that coefficient $h$ has little variation with $T_{\rm eff}$ or [M/H] in the interval $4.0 \leq \log T_{\rm eff} \leq 4.4$, staying within the range 0.5 to 0.4. This confirms a previous calculation of LDCs by D{\'{\i}}az-Cordov\'es et al.\ (1995), who found that the related square-root law works well at high $T_{\rm eff}$ values.

In this paper we have focused our attention on a comparison between the power-2 and quadratic laws, as the quadratic one is by far the most widely used bi-parametric law. Figures\ \ref{fig:4500k} and \ref{fig:23000k} illustrate the behaviour of the normalised specific intensity near the limb of the star as a function of $\mu$ for the models with $T_{\rm eff}$ values of 4500\,K and 23000\,K, respectively. The other input  parameters are the same for both models: $\log g = 5.0$, [M/H] $= 0.0,$ and $V_\xi = 2$\,km\,s$^{-1}$. In both cases the superiority of the power-2 law over the quadratic one is clear, mainly in the regions near the limb, where the differences in the normalised specific intensities can be of the order of 0.06. On the other hand, a comparison between Figs.\ \ref{fig:4500k} and \ref{fig:23000k} shows that the power-2 law works better for models with high $T_{\rm eff}$ values, mainly at the limb. We conducted tests with models with different input  parameters, and the results are similar, although they vary slightly in relation to the previously mentioned values, depending on the region of the Hertzsprung-Russell  diagram analysed.

A more general way of checking such differences in the quality of the fittings can be done by inspecting Figs.\ \ref{fig:pvq:gbp}--\ref{fig:pvq:tess}. These figures show the merit functions, ${\chi^2}$, for the space missions \textit{Gaia}, \textit{Kepler,} and TESS. A direct comparison shows that the power-2 law (red asterisks)  presents a much higher quality of fit than the corresponding quadratic one (black asterisks). However, we should recall that the two-parameter laws are accurate only for some regions of the Hertzsprung-Russell  diagram. For example, for atmospheres in radiative equilibrium, the power-2 law works very well. However, for models in convective equilibrium ($\log T_{\rm eff} \leq 3.85$), the power-2 law begins to present significant $\chi^2$ values, albeit still much lower than in the case of the quadratic law.


\section{Comparison with the square-root and logarithmic laws}

For completeness, we also compared the quality of fit for the power-2 law with those computed by adopting the square-root and logarithmic LD laws. These are rarely used in the analysis of transiting planets but are often used in the study of eclipsing binary systems (e.g.\ Van Hamme 1993).

Figure~\ref{fig:pvs:g} shows a comparison of the merit functions for the $G$ passband using the square-root law (black asterisks) and the power-2 (red asterisks). We have kept the same scale to facilitate a comparison with the results shown in Fig.~\ref{fig:pvq:g}.  It can be seen that the superiority of the fits provided by the power-2 law (Fig.~\ref{fig:pvs:g}) is not as great as for the quadratic law (Fig.~\ref{fig:pvq:g}) because the square-root law is a better representation than the quadratic law. In fact, D{\'{\i}}az-Cordov\'es \& Gim\'enez (1992) have shown that the values of $\sigma$ for the square-root law are of the order of ten times smaller than those provided by the quadratic one.  To assess the quality of a fit, D{\'{\i}}az-Cordov\'es \& Gim\'enez (1992) used a definition of the merit function slightly different from the one given by Eq. 7:

\begin{equation}
{\sigma^2} = {1\over{N}}\sum_{i=1}^{N} \left( {y_i - Y_i}\right)^2,
\end{equation}

\noindent
 where $N$, $y_i$, and $Y_i$ have the same meaning as in Eq. 7. 
 
Figure~\ref{fig:pvl:g} shows the same comparison but for the logarithmic law. It can be seen that the logarithmic law provides a better fit to the model atmosphere intensities than does the quadratic law  (Fig. 5), but it is clearly inferior to the square-root  (Fig. 9) and power-2 laws. This conclusion holds particularly at higher $T_{\rm eff}$ values, as the logarithmic law performs relatively well for $5500 < T_{\rm eff} < 9000$~K.  In summary, in addition to the fact that Eq.~5 is a non-linear law, as noted by Hestroffer (1997), it exhibits steeper $\partial I\over {\partial \mu}$ gradients at the limb (see, for example, Figs. 2 and 3). This makes it the bi-parametric law that best fits the distribution of intensities from atmosphere models, as we have seen by inspecting Figs.\ 4-10.


\section{Summary}

The power-2 LD law is seeing increasing use in several areas of stellar physics. In order to help this work, we have calculated LDCs for the power-2 law using ATLAS model atmospheres covering a wide range of stellar $T_{\rm eff}$ values, gravities, and metallicities. This has been done for a large number of passbands used by both ground-based and space-based telescopes.

 In addition to the LDCs, we have calculated the quality of fit to the specific intensities from the model atmospheres. A comparison between these and other two-parameter LD laws shows that the power-2 law produces the best results in all cases, followed by the square-root and logarithmic laws.  The quadratic law is by far the worst of the four options. The differences between the $\chi^2$ provided by the power-2 law and the quadratic one are particularly important for regions near the limb and for low $T_{\rm eff}$ values, and can reach more than one order of magnitude. We find similar results for comparisons in all of the passbands considered. Additional calculations for other photometric systems  and/or the transmissions of the filters and the response of the Earth's atmosphere are available upon request.

Therefore, we recommend the power-2 law to users who prefer to use a two-parameter LD law. We note that the four-parameter law provides a better fit, particularly in the cases of eclipsing binaries and extrasolar planet transits with stars that have low $T_{\rm eff}$ values. For the specific case of short-period eclipsing binaries, it would also be advisable to take the mutual irradiation into account because it can significantly alter both the corresponding LDCs and the bolometric albedo. In fact, irradiated stellar atmospheres show different distributions of brightness if we compare them with standard models without the action of an external radiation field. More details on this matter are available in Claret (2001, 2004) and Claret \& Gim\'enez (1992).


{}


\begin{acknowledgements}
 We thank the anonymous referee for his/her helpful comments
        that have improved the manuscript. The Spanish MEC (AYA2015-71718-R and
        ESP2017-87676-C5-2-R) is gratefully acknowledged for its support during the
The Spanish MEC (ESP2017-87676-C5-2-R,  PID2019-107061GB-C64, and PID2019-109522GB-C52) is gratefully acknowledged for its support during the development of this work. A.C.\ acknowledges financial support from the State Agency for Research of the Spanish MCIU through the "Center of Excellence Severo Ochoa'' award for the Instituto de Astrof{\'{\i}}sica de Andaluc{\'{\i}}a (SEV-2017-0709).  This research has made use of the SIMBAD database, operated at the CDS, Strasbourg, France, of NASA's Astrophysics Data System Abstract Service and of SVO Filter Profile  supported from the Spanish MINECO through grant AYA2017-84089.
\end{acknowledgements}


\renewcommand{\tablename}{Appendix A: Description of Tables 1-3 (available at the CDS)}
\begin{table*}

        \caption{Power-2 LDCs for the \textit{Gaia}, \textit{Kepler}, TESS, $ugriz$, $uvby,$ and $UBVRIJHK$ photometric systems.}
        \begin{flushleft}
                \begin{tabular}{lcccccclc}
                        \hline
                        Name & Source & $T_{\rm eff}$ values (K) & log $g$ (c.g.s.) & [M/H] (dex) & $V_\xi$ (km\,s$^{-1}$) & Photometric system & Fit/equation   \\
                        \hline
                        Table1 & {\sc ATLAS} & 3500\,--\,50000 & 0.0\,--\,5.0 & $-$5.0\,--\,0.0 & 0, 1, 2, 4, 8 &{\textit{Gaia}, \textit{Kepler}, TESS} & LSM/Eq. 5 \\
                        Table2 & {\sc ATLAS} & 3500\,--\,50000 & 0.0\,--\,5.0 & $-$5.0\,--\,0.0 & 0, 1, 2, 4, 8 &{$ugriz$}                              & LSM/Eq. 5 \\
                        Table3 & {\sc ATLAS} & 3500\,--\,50000 & 0.0\,--\,5.0 & $-$5.0\,--\,0.0 & 0, 1, 2, 4, 8 & $uvbyUBVRIJHK$                        & LSM/Eq. 5 \\
                        \hline
                \end{tabular}
        \end{flushleft}
\end{table*}

\end{document}